\documentclass[letter,twocolumn]{jpsj3} 
\usepackage{bm} 

\newcommand{\degree}{{$^\circ$}}

\title{Crossover from commensurate to incommensurate antiferromagnetism in stoichiometric
NaFeAs revealed by single-crystal $^{23}$Na,$^{75}$As-NMR experiments}

\author{Kentaro~\textsc{Kitagawa$^{1,2}$}\thanks{kitag@issp.u-tokyo.ac.jp},
Yuji~\textsc{Mezaki$^{1,3}$},
Kazuyuki~\textsc{Matsubayashi$^{1,2}$},
Yoshiya~\textsc{Uwatoko$^{1,2}$},
and Masashi~\textsc{Takigawa$^{1,2}$}}

\inst{$^1$Institute for Solid State Physics, University of Tokyo,
5-1-5 Kashiwanoha, Kashiwa, Chiba 277-8581, Japan\\
$^2$JST, TRIP, 5 Sanbancho, Chiyoda, Tokyo 102-0075, Japan\\
$^3$College of Science and Technology, Nihon University, Kanda-Surugadai 1-8, Chiyoda-ku, Tokyo 101-8308, Japan\\}

\abst{We report results of $^{23}$Na and $^{75}$As nuclear magnetic 
resonance (NMR) experiments on a self-flux grown high-quality single 
crystal of stoichiometric NaFeAs. The NMR spectra revealed a 
tetragonal to twinned-orthorhombic structural phase transition at 
$T_\text{O}$ = 57~K and an antiferromagnetic (AF) transition 
at $T_\text{AF}$ = 45~K.  The divergent behavior of nuclear 
relaxation rate near $T_\text{AF}$ shows significant anisotropy, indicating 
that the critical slowing down of stripe-type AF fluctuations are strongly 
anisotropic in spin space. The NMR spectra at low enough temperatures 
consist of sharp peaks showing a commensurate stripe AF order with a 
small moment $\sim$0.3~$\mu$B. However, the spectra just below 
$T_\text{AF}$ exhibits highly asymmetric broadening pointing to an 
incommensurate modulation. The commensurate-incommensurate 
crossover in NaFeAs shows a certain similarity to the behavior of 
SrFe$_2$As$_2$ under high pressure.}
\kword{NaFeAs, NMR, incommensurate antiferromagnetism, 111-type iron arsenide}

\begin{document}
\maketitle
Stripe-type antiferromagnetic (AF) order has been commonly observed 
on the undoped or low-pressure region in the phase diagrams of iron-based
superconductivities.
Thus AF fluctuations are often considered the driving force for 
superconducting (SC) paring\cite{IshidaJPSJReview}. For example, 
the $s_\pm$-wave pairing state, where the order parameter changes 
sign between the nested electron and hole Fermi surfaces, can be stabilized 
by the interband AF fluctuations\cite{Mazinsplusminus}. Since the geometry of the 
Fermi surfaces depends on materials and changes with doping, the nesting wave vectors may
deviate from the commensurate (C) conditions in some cases, leading to incommensurate (IC)-AF order, or in
other words, spin-density-wave (SDW) order in a narrow sense.

However, all the AF orders of the undoped stoichiometric compounds so far are known to be 
C-type\cite{LumsdenMagnetismReview}.  IC order has been 
reported for some non-stoichiometric or chemically doped materials.  
Neutron scattering experiments on the lightly electron-doped systems
Fe$_{1+\delta}$Te$_{1-x}$Se$_x$ reported short-ranged IC order 
characterized by broad magnetic peaks off-centered from the 
commensurate wave vectors\cite{WenFeTeSeNS,BaoFeTeSeNS}. 
IC-SDW order has been also reported for 
Ba(Fe$_{1-x}$Co$_x$)$_2$As$_2$ by NMR\cite{LaplaceBa122CoNMR} 
and M{\"o}ssbauer\cite{BonvilleBa122CoMB} experiments.       
However, microscopic disorder in these materials should bring 
distribution in the magnitude of ordered moments, which might be difficult to 
distinguish from incommensuration due to Fermi surface nesting. 
Therefore, finding a stoichiometric material exhibiting IC-SDW 
would mark important advance in our understanding of magnetism 
in iron-pnictide materials.   

\begin{figure}[t]
\centering
\includegraphics[width=1.0\linewidth]{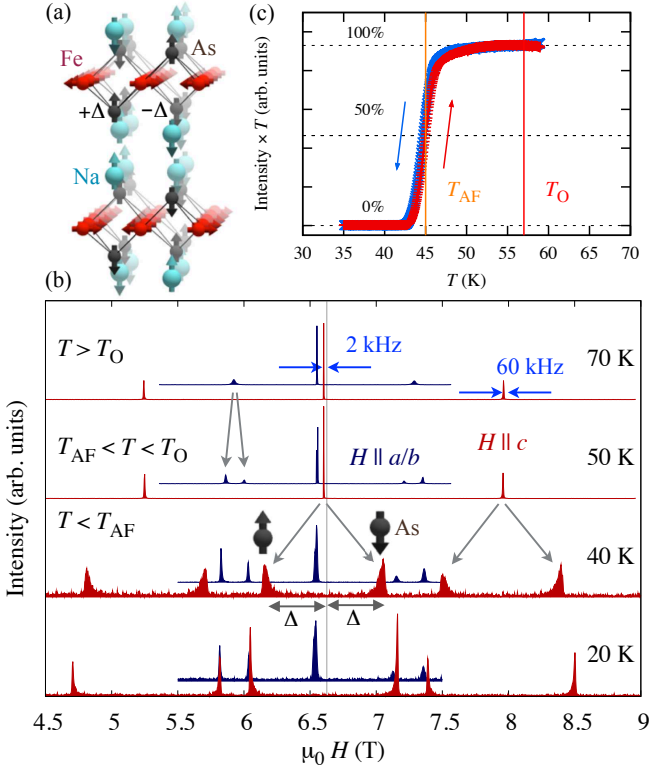}
\caption{(Color online) (a) Crystal structure and stripe-type
antiferromagnetic order (red arrows) of NaFeAs. The arrows 
at As and Na nuclei indicate the hyperfine field in the antiferromagnetic state. 
(b) Field-swept $^{75}$As-NMR spectra at 48.31~MHz. 
Below $T_\text{O}$, the satellite lines for $H \perp c$ splits into two sets due to twining.
Below $T_\text{AF}$, the center line for $H \parallel c$ splits
symmetrically by the hyperfine fields parallel to the $c$ axis at the As sites. 
The dotted line shows the unshifted resonance field.
The full widths at the half maxima (enclosed by the arrows) are narrower than those 
for 122 parent compounds\cite{KitagawaBa122,KitagawaSr122}, indicating good 
stoichiometry of the sample. (c) Temperature dependence of the intensity 
at the peak position of the paramagnetic spectrum multiplied by $T$ for 
$H \parallel c$ . No hysteresis was observed across $T_\text{AF}$,
except for the small instrumetal delays.}
\label{fig:spectra75}
\end{figure}

Among different types of iron-based superconductors, 122-type 
$A$Fe$_2$As$_2$ ($A$=Ca, Sr, Ba) have been attracting strong 
interest because large single crystals with stoichiometric composition 
can be grown.  Recently 111-type materials, LiFeAs and NaFeAs 
(Fig.~\ref{fig:spectra75}a), have also been found suitable for crystal growth 
with alkaline-rich self flux\cite{ChenNaFeAs,BorisenkoLiFeAs}. 
LiFeAs is a bulk superconductor with the transition temperature $T_\text{SC}$ of 
17~K\cite{TappLiFeAs}, while NaFeAs shows only filamentary
superconductivity\cite{YuNaFeAsNMR,LiNaFeAsNS}.
NaFeAs shows successive phase transitions: tetragonal-to-orthorhombic 
structural transition at $T_\text{O} \sim$ 50~K and AF transition at
$T_\text{AF} \sim$ 40~K\cite{ChenNaFeAs,LiNaFeAsNS}.  The 
transition temperatures, which are significantly lower than those for 1111 
and 122 materials\cite{LumsdenMagnetismReview}, seem to depend 
sensitively on self-doping due to Na deficiency or synthetic 
methods\cite{LiNaFeAsNS}.   In this letter, we report observation 
of IC-AF order and a temperature-driven IC-C crossover in 
stoichiometric NaFeAs by NMR experiments on $^{23}$Na and $^{75}$As nuclei. 

The crystals of NaFeAs were 
grown by the self-flux method with starting elemental ratio of 2:1:2.  
The materials were put into alumina crucible, sealed in a
doubled-wall quartz tube in argon atmosphere, and heated to
900\degree C.  Crystals grew during subsequent cooling 
down to 400\degree C in three days, then to room temperature 
in one day. We confirmed 1:1:1 atomic content within 2\% by energy 
dispersive X-ray spectroscopy.  Resistivity, magnetization 
and specific heat measurements detected the successive transitions 
at $T_\text{O}$ = 57~K and $T_\text{AF}$ = 45~K.  The values of 
$T_\text{O}$ and $T_\text{AF}$ of our crystal, also 
confirmed by the NMR experiments as described below, are significantly 
higher than the previous reports\cite{YuNaFeAsNMR,LiNaFeAsNS},
indicating less self-doping. The $c$-axis length measured by 
single-crystal x-ray diffraction was 7.05~\AA\ at room temperature, 
which is longer than the previous reports.  The crystal 
($5\times 3\times 0.3$~mm$^3$) was handled in purified argon and 
covered with grease before exposed to air for mounting 
on the NMR probe equipped with a double-axis goniometer. 
The NMR spectra were acquired using Fourier step summing technique. 
The Fourier transformed spectrum of spin-echo was accumulated 
at shifted frequencies while the frequency or magnetic field was being swept.
NMR relaxation rates $T_1^{-1}$ were determined by the 
inversion recovery method. Good fitting to the theoretical 
recovery curve for spin $I=3/2$ nuclei, 
$\{ \exp(-t/T_1) + 9 \exp(-6t/T_1)\}/10$, was obtained in the whole temperature range.

\begin{figure}[th]
\centering
\includegraphics[width=1.0\linewidth]{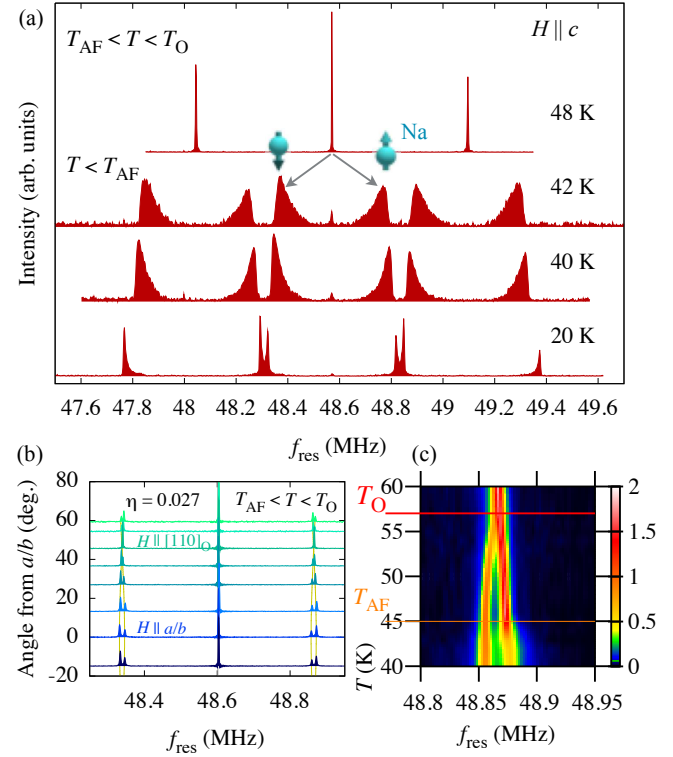}
\caption{(Color online) Frequency-swept $^{23}$Na-NMR spectra at 4.25~T (a) for 
$H \parallel c$ at different temperatures and (b) for $H \perp c$ along various 
directions in the $ab$-plane at $T$ = 50~K. The spectra for $H \parallel c$ 
show incommensurate modulation immediately below $T_\text{AF}$. 
The satellite lines are doubled for $H \perp c$ due to twinning in the 
orthorhombic structure. Their angular dependence in the $ab$-plane is 
well reproduced by eq.~\eqref{eq:nuq} with $\eta$ = 0.027 (dashed lines). 
(c) A color plot of the satellite intensity as a function of temperature 
for $H \parallel a$ or $b$, showing the tetragonal-to-orthorhombic 
structural transition at $T_\text{O}$ = 57~K.}
\label{fig:spectra23}
\end{figure}
The NMR spectra are shown in Fig.~\ref{fig:spectra75}(b) for $^{75}$As 
nuclei and in Fig.~\ref{fig:spectra23} for $^{23}$Na nuclei.  Both As and
Na occupy the same type of site, the 2$c$ site in the high-$T$ tetragonal 
$P4/nmm$ structure and the 4$g$ site in the low-$T$ orthorhombic $Cmma$ 
structure. Since both nulcei have spin $3/2$, the NMR spectra consist of 
quadrupole-split three lines with the resonance frequencies
(here up to the first-order perturbation of the quadrupole interaction).
\begin{equation}\label{eq:resonance}
f_\text{res}= \mu_0 \gamma_\text{N}H_\text{eff} + i\delta\nu \ \  (i=-1, 0, 1), 
\end{equation}
where $\gamma_\text{N}$ is the nuclear gyromagnetic ratio 
($2\pi \times 7.29019$~MHz/T for $^{75}$As and $2\pi \times 11.26226$~MHz/T 
for $^{23}$Na), $\bm H_\text{eff} = \bm H_\text{ext} + \bm H_\text{hf}$ is the sum 
of the external field and the magnetic hyperfine field, and 
\begin{equation}\label{eq:nuq}
\delta\nu = \frac{\nu^c}{2}\left(3\cos^2\theta - 1 + \eta\sin^2\theta \cos 2\phi\right),
\end{equation}
is the quadrupole splitting. Here, $(\theta,\phi)$ specifies the polar angle of 
$\bm H_\text{eff}$ with respect to the $c$-axes and  
$\eta = |\nu^{aa} - \nu^{bb}| / |\nu^{cc}|$ is the asymmetry parameter of the 
electric field gradient (EFG) tensor $\nu_{\alpha \alpha}$.  In the tetragonal 
$P4/nmm$ structure, $\eta$ must be zero. While the center line ($i$ = 0) is determined 
only by the magnetic hyperfine field, therefore, provides direct information on the spin structure, 
the satellite lines ($i = \pm 1$) are affected by EFG which is a sensitive probe for structural 
transitions. 

The $^{75}$As-NMR spectrum at 70~K (Fig.~\ref{fig:spectra75}b) shows 
three lines for both $H \parallel c$ and $H \perp c$ as expected. 
At 50~K below $T_\text{O}$ the satellite lines are doubled for 
$H \parallel a$ or $H \parallel b$ in the orthorhombic notation. 
This has been observed also in 122-type compounds and ascribed 
to non-zero $\eta$ and twinned domains in 
the orthorhombic structure\cite{KitagawaBa122,KitagawaSr122}.    
Details of the transition were investigated by $^{23}$Na NMR (Fig.~\ref{fig:spectra23}) providing better S/N ratio.
The angular dependence of the satellite resonance frequencies 
in the $ab$-plane at 50~K (Fig.~\ref{fig:spectra23}b) is well reproduced 
by Eq.~\eqref{eq:nuq} with $\eta$ = 0.027 (dashed lines). The onset 
temperature for the splitting of the satellite lines (Fig.~\ref{fig:spectra23}c) 
agrees with $T_\text{O}$ = 57~K determined by the resistivity measurement.  

The AF transition is marked by vanishing of the 
NMR line for $H \parallel c$ near the paramagnetic 
resonance field below  $T_\text{AF}$ = 45~K (Fig.~\ref{fig:spectra75}c for $^{75}$As). 
Absence of hysteresis indicates the transition is continuous. 
At sufficiently low temperatures, each of three lines split 
symmetrically into two lines for $H \parallel c$, but no line splitting 
occurs for $H \perp c$ [see the spectra at $T = 20$~K for $^{75}$As 
(Fig.~\ref{fig:spectra75}b) and $^{23}$Na (Fig.~\ref{fig:spectra23}a)].
This indicates 
staggered hyperfine fields at the As/Na sites along the 
$c$-direction $\bm H_\text{hf} = (0,0,\pm\Delta)$ with very small distribution of $\Delta$.
Following the analysis on 
the similar results for 122 compounds\cite{KitagawaBa122, KitagawaSr122}, 
we conclude that the hyperfine field is generated by a stripe AF 
order with the $q$-vector $(1,0,\frac{1}{2})_\text{O}$ (defined on 
the orthorhombic $Cmna$ unit cell) and the AF moment along the 
$a$-axis (Fig.~\ref{fig:spectra75}a). The off-diagonal element 
($B_{ac}$) of the hyperfine coupling tensor relates the AF moment 
$\sigma$ and $\Delta$ as $\Delta=4B_{ac}\sigma$\cite{KitagawaBa122}.
Assuming the same value $B_{ac}=0.4 \sim 0.5$~T/$\mu_\text{B}$ obtained 
for 122 compounds\cite{KitagawaBa122, KitagawaSr122}, 
the AF moment in NaFeAs is estimated to be 
$0.28 \sim 0.34$~$\mu_\text{B}$ at 20~K.  
This is close to the value reported by Yu {\it et\, al.} also from NMR 
experiments\cite{YuNaFeAsNMR} but significantly larger than the value 
$0.09 \pm 0.04$~$\mu_\text{B}$ obtained by neutron experiments\cite{LiNaFeAsNS}. 

However, the spectral shape exhibits unusual temperature dependence near $T_\text{AF}$.
The narrow six lines of the 
$^{75}$As NMR spectrum at 20~K develop asymmetric distribution at 40~K 
with a sharp edge at the maximum value of $\Delta$ (Fig.~\ref{fig:spectra75}b). 
The sharp upper cutoff is unlikely to be caused by extrinsic disorder or inhomogeneity.
This spectral broadening is more obvious in the $^{23}$Na spectra immediately below 
$T_\text{AF}$ (Fig.~\ref{fig:spectra23}a), indicating incommensurate 
modulation of the hyperfine field.  However, the spectral shape changes 
gradually to rather normal discrete peaks at lower temperature, pointing to 
an incommensurate-commensurate crossover.  Since the broadening is 
observed only for $H \parallel c$, the hyperfine fields at the As/Na 
sites should still be almost parallel to the $c$ axis, $\bm H_\text{hf} = (0,0,\pm\Delta)$.  
The spectral shape then represents the spatial distribution of $\Delta$ or the 
AF moments.
Therefore, the NMR spectra near $T_\text{AF}$ indicate a longitudinal SDW order at
some incommensurate wave vector.

\begin{figure}[htb]
\centering
\includegraphics[width=1.0\linewidth]{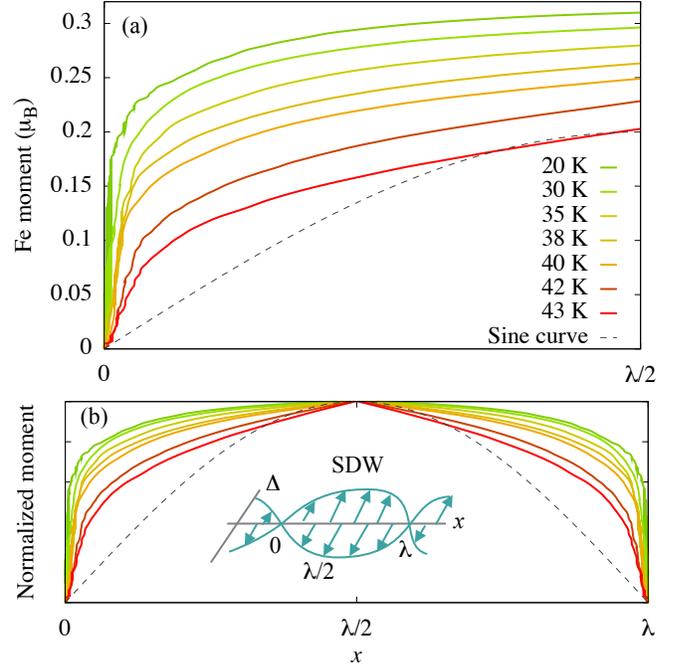}
\caption{(Color online) Spatial distribution of Fe moments in the AF state of NaFeAs,
assuming the simple periodicity (see text). The panel (a)
corresponds to the region $0 < \Theta < \pi/2$ in the schematic in (b).
(b) The temperature dependence of the SDW patterns, illustrated by normalizing the data in (a).}
\label{fig:sdw}
\end{figure}
To illustrate the real-space image of moment distribution, 
we assume a periodic modulation of $\Delta$ between zero and 
$\Delta_\text{max}$ along $x$ over the wave length $\lambda$.  
The spectral intensity for $H \parallel c$ as a function of $\Delta$ 
is given as 
\begin{equation}
I(\pm\Delta) \propto \left|\frac {d \Delta (x)}{d x}\right|^{-1}.
\end{equation}
Integrating the experimental spectra, therefore, gives the relation between $\Delta$ and $x$, 
\begin{equation}
x = \frac{\lambda}{2} \frac{\int^\Delta_0 I(\Delta^{\prime}) d\Delta^{\prime}}{\int^{\Delta_\text{max}}_0 I(\Delta^{\prime}) d \Delta^{\prime}}.
\end{equation}
By converting $\Delta$ to the AF moment, the spatial distribution of AF
moments reproduced by $^{23}$Na-NMR
 is obtained as shown in Fig.~\ref{fig:sdw} at various temperatures. At low
temperatures, the AF moment is nearly uniform over most of the space and the
rapid change occurs only in a narrow region. Such a profile can be regarded
as usual C-AF domains separated by domain walls. Near $T_\text{AF}$, however, the
domain wall gets significantly broadened, making the profile closer to a
sinusoidal SDW. The microscopic mechanism for such IC-C crossover is still
an open question.
 
The IC-SDW order at $(1 + \delta,0,\frac{1}{2})_\text{O}$ or 
$(1, \delta,\frac{1}{2})_\text{O}$ breaks cancellation of the $a$-component 
of the hyperfine field\cite{KitagawaBa122}, therefore, 
should broaden the NMR spectra for $H \perp c$.  The full-width of the 
spectrum for $H \perp c$ due to incommesuration is given by 
$\sim 8B_\text{aa}\sigma\tan (\pi \delta)$, where $B_{aa}$($\simeq {^{75}A^{ab}}/4$) is the 
diagonal element of the hyperfine coupling tensor determined from the Knight shifts described below.
The observed width (0.02~T at 43~K) then puts an upper 
limit for $\delta$,  $|\delta| < 0.004$.  Such tiny deviation from commensuration 
might be difficult to detect by diffraction experiments. Similar observation was 
reported for Co-doped BaFe$_2$As$_2$\cite{LaplaceBa122CoNMR}.  
We cannot rule out, however, interlayer modulation 
$(1, 0, \frac{1}{2}+\delta)_\text{O}$ with larger $\delta$.

We note that an IC-C crossover has been observed in 
SrFe$_2$As$_2$ under high pressure near 5.4~GPa, where AF and SC states 
coexist, most likely  forming nano-scale hybrid structure\cite{KitagawaSrFe2As2UHP}. 
The $^{75}$As NMR spectra from AF region show commensurate peaks 
at low-$T$ but broad distribution at high-$T$ with a crossover near 18~K. Thus the 
incommensurate stripe order may not be exceptional but common in iron-based 
materials, in particular, in the vicinity of superconducting phases when $T_\text{AF}$ 
is depressed below 50~K.
 
\begin{figure}[htb]
\centering
\includegraphics[width=1.0\linewidth]{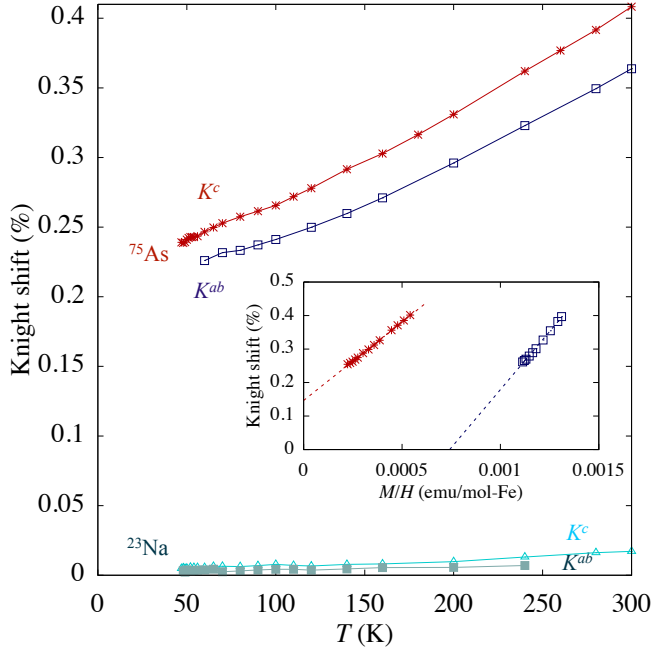}
\caption{(Color online) Knight shifts in the paramagnetic state.
The inset shows the $K$ versus $\chi$ plots for the As sites (low-$T$ data below 70~K
 have been omitted since $M/H$ showed upturns caused by impurities).}
\label{fig:kvst}
\end{figure}

We next discuss the results in the paramagnetic state.  
The Knight shifts $K$ ($=H_\text{eff}/H$) at the $^{75}$As and $^{23}$Na
nuclei are plotted against temperature in Fig.~\ref{fig:kvst}
after correcting for the second order quadrupole effects.
The Knight shifts at the Na sites are small, of the order of the dipolar 
field from Fe moments. By plotting $K$ at the As sites against the 
susceptibility (the inset of fig.~\ref{fig:kvst}), we obtain the diagonal 
elements of the hyperfine coupling tensor: ${^{75}A^c} = 2.61 \pm 0.02$~T/$\mu_\text{B}$, 
$^{75}A^{ab} = 3.87 \pm 0.07$~T/$\mu_\text{B}$.

\begin{figure}[tbh]
\centering
\includegraphics[width=1.0\linewidth]{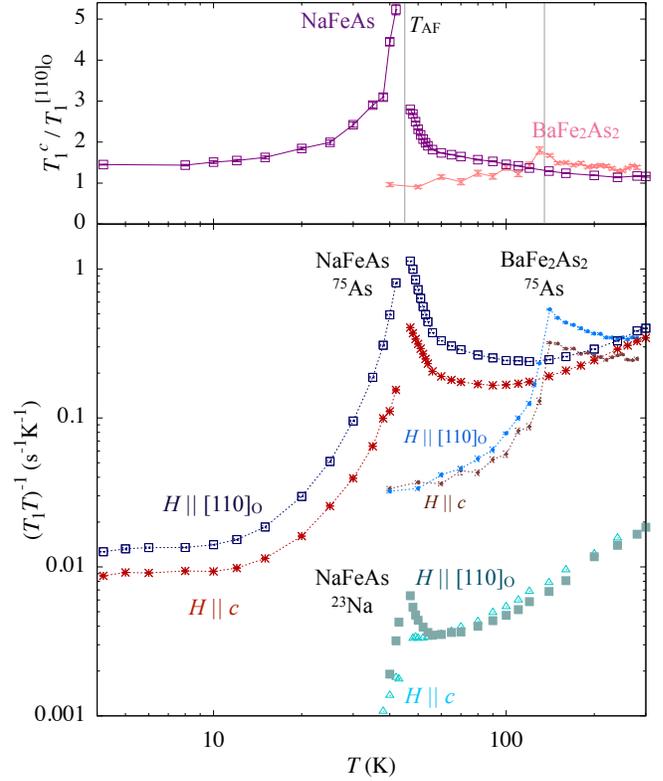}
\caption{(Color online) The lower panel shows the spin-lattice relaxation 
rate divided by temperature $(T_1T)^{-1}$ in 
NaFeAs, and BaFeAs\cite{KitagawaBa122} for two directions of the field. 
Upper panel shows the anisotropy $T_1^c / T_1^{[110]}$ of the $^{75}$As-NMR.}
\label{fig:t1t}
\end{figure}
The lower panel of Fig.~\ref{fig:t1t} shows the temperature 
dependence of the relaxation rate divided by $T$. The Korringa 
law, $(T_1T)^{-1} \sim const.$, is observed at the lowest temperatures 
both in NaFeAs and in 122 compounds\cite{KitagawaBa122,KitagawaSr122}, 
indicating that a part of the Fermi surface survives in the AF state 
even in the presence of a SDW gap. 
The divergence of $(T_1T)^{-1}$ in NaFeAs near $T_\text{AF}$ is 
compatible with a second order transition. Then the nuclear relaxation 
should be dominated by the stripe AF fluctuations near the wave 
vector $\bm Q=(1,0,\frac{1}{2})_\text{O}$.  
The anisotropic ratio of the relaxation rate
${(T_1^{\text{[110]}})^{-1}}/{ (T_1^{c})^{-1} }$ provides a measure 
of anisotropy of the AF fluctuations in spin space\cite{KitagawaSr122}, 
\begin{equation}\label{eq:t1aniso}
\frac{(T_1^{\text{[110]}})^{-1}}{ (T_1^{c})^{-1} } = 
\frac{
2|S^{a}(\bm Q, \omega_\text{res})|^2
 +|S^{c}(\bm Q, \omega_\text{res})|^2}
{2|S^{c}(\bm Q, \omega_\text{res})|^2},
\end{equation}
where $|S^{i}(\bm Q, \omega_\text{res})|^2$ is the power spectrum of the spin fluctuation
along the $i$-direction at $\bm Q$ and the NMR frequency $\omega_\text{res}$. 
The upper panel of Fig.~\ref{fig:t1t} shows divergent behavior of the ratio
${(T_1^{\text{[110]}})^{-1}}/{ (T_1^{c})^{-1} }$ of $^{75}$As-NMR, which translate 
into the development of strong anisotropy of the AF fluctuations
$|S^{a}(\bm Q, \omega_\text{res})| \gg |S^{c}(\bm Q, \omega_\text{res})|$. 
In BaFe$_2$As$_2$ the divergence is interrupted by the first order transition. 

In summary, we have investigated the structural and magnetic transitions and 
AF spin structure in NaFeAs using $^{75}$As- and $^{23}$Na-NMR.
The NMR relaxation data prove the second-order nature of the AF
transition and the strong anisotropy of the AF fluctuations in spin space. 
The most remarkable feature of the AF state is the temperature-driven 
crossover from the high-$T$ incommensurate structure to the low-$T$ 
commensurate structure.  This must be an intrinsic feature of a clean and 
stoichiometric system, as demonstrated by the very sharp NMR spectra 
and phase transitions of our crystal.

We thank N.~Katayama, Y.~Kiuchi, and the Materials Design and Characterization Laboratory in ISSP for experimental supports.
This work was supported partly by the Grant-in-Aid for Scientific Research (B) (No. 21340093) from
JSPS and by the GCOE program from MEXT Japan.

\bibliography{document}

\end{document}